\documentclass[12pt,nohyper,letterpaper]{JHEP3}         
\usepackage{amssymb,amsfonts}

\usepackage{epsfig}

\def\be{\begin{eqnarray}}
\def\ee{\end{eqnarray}}
\newcommand{\nn}{\nonumber}
\newcommand\para{\paragraph{}}
\newcommand{\ft}[2]{{\textstyle\frac{#1}{#2}}}
\newcommand{\eqn}[1]{(\ref{#1})}

\newcommand\valpha{\vec{\alpha}}
\newcommand\vg{\vec{g}}

\def\Dslash{\,\,{\raise.15ex\hbox{/}\mkern-12mu D}}
\def\Dbarslash{\,\,{\raise.15ex\hbox{/}\mkern-12mu {\bar D}}}
\def\delslash{\,\,{\raise.15ex\hbox{/}\mkern-9mu \partial}}
\def\delbarslash{\,\,{\raise.15ex\hbox{/}\mkern-9mu {\bar\partial}}}
\def\pslash{\,\,{\raise.15ex\hbox{/}\mkern-9mu p}}
\def\calDslash{\,\,{\raise.15ex\hbox{/}\mkern-12mu {\cal D}}}

\newcommand{\CP}{{\bf CP}}

\newcommand{\Tr}{{\rm Tr}}

\def\lae{\mathrel{\mathop{\smash{\lower .5 ex \hbox{$\stackrel<\sim$}}}}}
\def\lae{\mathrel{\mathop{\smash{\lower .5 ex \hbox{$\stackrel>\sim$}}}}}

\title{D-Branes in Field Theory}
\author{David Tong \\
Department of Applied Mathematics and Theoretical Physics, \\
University of Cambridge, UK\\
{\tt d.tong@damtp.cam.ac.uk}}

\abstract{Certain gauge theories in four dimensions are known to
admit semi-classical D-brane solitons. These are domain walls on which vortex flux
tubes may end. The purpose of this paper is to develop an
open-string description of these D-branes. The dynamics of the
domain walls is shown to be governed by a Chern-Simons-Higgs
theory which, at the quantum level, captures the classical
scattering of domain wall solitons.}

\begin{document}
\pagestyle{plain} \setcounter{page}{1}
\newcounter{bean}
\baselineskip16pt

\section{Introduction}

Ten years ago, Polchinski introduced the concept of a D-brane
\cite{pol}. It is hard to overstate the role that these objects
have played in theoretical physics in the intervening time. They
have lead to increased understanding of dualities, given impetus
to creative new ideas in phenomenology and cosmology and, perhaps
most importantly, unpin our understanding of strongly coupled gauge
theory through the AdS/CFT correspondence.

\para
The purpose of this paper is to explore the dynamics of D-branes
in the rather more mundane setting of field theory, decoupled from
gravity and the associated complications. In particular, we will
determine the ``open-string" description of the dynamics of
D-branes in four dimensional ${\cal N}=2$ super QCD. Part of the
motivation for this work is to study the string-gauge theory
correspondence for gauge theories in the Higgs phase where the
strings in question are semi-classical magnetic flux tubes whose
worldvolume dynamics is under good control.

\para
A D-brane is defined as a hypersurface on which a string may end.
In a field theory, both the brane and the string itself must
arise as solitonic type objects. With this definition there are
several systems appearing in Nature which can be said to admit
D-branes, including superfluid ${}^3$He and anti-ferromagnets.
However, we may ask for two further requirements from our D-branes
so that they are closer in spirit to those appearing in string
theory. The first is that the brane houses a $U(1)$ gauge field
living on its worldvolume under which the end of the string is
charged. The second requirement is that as two D-branes approach,
the stretched string gives rise to a new light excitation on its
worldvolume which governs the dynamics of the brane. There are at least three
examples of non-gravitational field theories admitting D-branes in
this stronger sense. In each case, the D-branes are D2-branes:
\begin{itemize}
\item ${\cal N}=1$ super Yang-Mills theory in four dimensions. The
theory lies in the confining phase. There exist  BPS domain walls
which are D-branes for the QCD flux tube \cite{witten}.
\item ${\cal N}=2$ super QCD in four dimensions. The theory lies
in the Higgs phase. Again there exist BPS domain walls which, this
time, are D-branes for the solitonic magnetic vortex
\cite{dbrane,sy1}.
\item ${\cal N}=(1,1)$ super Yang-Mills in six dimensions, which
can be thought of as the low-energy limit of type iib little
string theory. The theory lies in the Coulomb phase. The spectrum
of solitons includes an instanton string and a monopole 2-brane.
The latter is a D-brane for the former \cite{hw}.
\end{itemize}
In string theory there are two methods to determine the dynamics
of D-branes. The ``closed string" description --- which in
practice means the supergravity approximation ---  is analogous to
the method used to study soliton scattering in field theory; one
simply follows the evolution of the classical bulk configurations under
the equations of motion. For our three examples above, this
procedure can be carried out explicitly, using the moduli
space approximation, only for the second and third cases where the
field theory is weakly coupled and the solitons semi-classical.

\para
The second, dual, method that exists in string theory is the
``open string" description. Here one neglects the back-reaction of
the D-branes on the spacetime fields, concentrating instead on the
light states on the D-brane worldvolume. As two D-branes approach,
new light states appear arising from string stretched between
them. For solitons in field theories, such an open-string
description does not usually exist. However, when the solitons are
D-branes for some solitonic string, one may wonder whether it is
possible to formulate an open string description of the dynamics.
For monopoles in six-dimensions, such a description was given in
\cite{hw} and for domain walls in ${\cal N}=1$ super Yang-Mills
an open string description was suggested in \cite{bobby}. We
review both of these theories in section 5.

\para
The goal of this paper is to derive an open string description for
the dynamics of domain walls in ${\cal N}=2$ SQCD\footnote{A
review of these domain walls, and their relationships with other
solitons, can be found in \cite{tasi}.}. We will show that the
dynamics is captured by a Chern-Simons-Higgs theory living on the
worldvolume of the walls.

\para
The plan of this paper is as follows. In the next section we will
quickly review some classical aspects of the four dimensional
abelian theory and show how the domain wall is a D-brane for the
vortex string. In section 3 we turn to domain wall interactions.
After reviewing the ``closed-string" moduli space approximation to
domain wall dynamics, we show how one can re-derive the results
using open-strings. Section 4 deals with domain walls in the
non-abelian theory. We end, in section 5, with a discussion.

\section{Domain Walls as D-Branes}

In this section and the next we start by studying the
the simplest D-branes in abelian models. The theory is
${\cal N}=2$ SQED, consisting of of $U(1)$ vector multiplet and
$N_f$ charged hypermultiplets. To describe the soliton solutions
we will require only a subset of the fields: the gauge field, a
real neutral scalar $\phi$ and $N_f$ charged scalars $q_i$. The
scalar potential is dictated by supersymmetry,
\be V=\sum_{i=1}^{N_f} (\phi-m_i)^2|q_i|^2+\frac{e^2}{2}
(\sum_{i=1}^{N_f} |q_i|^2-v^2)^2\label{lag}\ee
%
The model depends on several parameters. The gauge coupling $e^2$
sits in front of the D-term in the potential. In addition, there
are real masses $m_i$ for each flavor and a Fayet-Iliopoulos (FI)
parameter $v^2$ which induces a vacuum expectation value (vev) for
$q_i$. When the theory has vanishing masses it enjoys an $SU(N_f)$
flavor symmetry with the $q_i$ transforming in the fundamental
representation. Distinct masses explicitly break this to the
maximal torus,
\be SU(N_f)\ \stackrel{m_i}{\longrightarrow}\
U(1)_F^{N_f-1}\label{u1f}\ee
which acts by rotating the phases of the $q_i$ individually,
modulo the $U(1)$ gauge group. The theory has $N_f$ isolated vacua
$V=0$ given by
\be \phi=m_i\ \ \ ,\ \ \ |q_j|^2=v^2\delta_{ij}\ \ \ \ \ \ \ {\rm
for}\ i=1,\ldots, N_f\ee
In each vacuum the $U(1)$ gauge symmetry is spontaneously broken
and the theory exhibits a gap. The photon has mass $M_\gamma=ev$,
while the remaining $N_f-1$ scalars that are not eaten by the
Higgs mechanism have masses $M_q=|m_j-m_i|$ with $j\neq i$.

\para
Classically we may take the strong coupling limit
$e^2\rightarrow\infty$ of the theory. This decouples the photon,
leaving behind $N_f-1$ complex degrees of freedom. For vanishing
masses, these degrees of freedom give rise to the sigma model with
target space\footnote{If we keep only half the hypermultiplet scalars
$q_i$ as in \eqn{lag} and neglect the complex scalars with opposite
charge (usually denoted $\tilde{q}_i$) then the target space is the
zero section ${\bf CP}^{N_f-1}$. Classically it is consistent to
restrict attention to the massive ${\bf CP}^{N_f-1}$ sigma-model since
only these modes are excited in the D-brane solitons. However,
in the quantum theory, fermions induce an anomaly in the
restricted $\CP^{N_f-1}$ theory \cite{sigmanom}, and we must
consider the full $T^\star \CP^{N_f-1}$ target space.} $T^\star\CP^{N_F-1}$. This
arises from the classical Lagrangian through the standard ``gauged
linear sigma model" technique \cite{phases}, in which the hypermultiplet
fields are constrained by the D-term, subject to gauge identification.
Non-zero masses $m_i$ induce a
potential on this target space with $N_f$ zeroes. This potential
can be shown to be proportional to the length of an appropriate
Killing vector on the target space, associated to the flavor
symmetry.

\para
At the quantum level, the four-dimensional sigma model is
non-renormalizable. Although we only require classical properties
of this theory, it is an implicit assumption that a suitable UV
completion exists. We shall comment further on this in section 5.

\para
The existence of isolated, gapped vacua ensures the existence of
domain walls in the theory. Consider a domain wall which
interpolates between vacuum $\phi=m_i$ as $x^3\rightarrow -\infty$
and $\phi=m_j$ as $x^3\rightarrow +\infty$. We choose to order the
masses as $m_i<m_{i+1}$. Then, for $i<j$, the first order
Bogomolnyi type equations describing the domain wall are
\be
\partial_3\phi=-\frac{e^2}{2}(\sum_{i=1}^{N_f}|q_i|^2-v^2)\ \
\ \ ,\ \ \ \ {\cal D}_3q_i=-(\phi-m_i)q_i\label{dw}\ee
subject to the appropriate boundary conditions as $x^3\rightarrow
\pm \infty$. Solutions to these equations describe $1/2$-BPS
domain walls with tension
\be T_{\rm wall}=v^2(m_j-m_i) \ee
The theory also admits vortex strings. In the $i^{\rm th}$ vacuum,
these are supported by the phase of scalar $q_i$ winding in the
plane transverse to the string. The first order equations
describing a string lying in the $x^3$ direction are
\be B_3=e^2(|q_i|^2-v^2)\ \ \ ,\ \ \  \ {\cal D}_1q_i=i{\cal
D}_2q_i\label{vortex}\ee
Solutions to these equations with unit winding number describe
$1/2$-BPS vortex strings. They have width $L_{\rm vortex}\sim
1/ev$ and tension,
\be T_{\rm vortex} = 2\pi v^2 \ee
Note that in the limit $e^2\rightarrow\infty$, these classical
strings become infinitesimally thin.

\subsection{The Single Wall as a D-Brane}

Let's review some properties of the domain wall in the simplest
$N_f=2$ theory with two vacua. We will show that this wall houses
a $U(1)$ gauge field and plays the role of a D-brane for the
vortex string \cite{dbrane,sy1}

\para
The single wall has two collective coordinates, both
Nambu-Goldsone modes \cite{at}. The first is simply the center of
mass of the domain wall, $X$, arising from broken translational
invariance. The second comes from the $U(1)_F$ flavor symmetry
\eqn{u1f},
\be U(1)_F: \ \ \ q_1\rightarrow e^{i\theta}q_1\ \ \ {\rm and}\ \
\ q_2\rightarrow e^{-i\theta}q_2\ee
In each vacuum either $q_1$ or $q_2$ vanishes and $U(1)_F$
coincides with the gauge action. However in the core of the domain
wall, both $q_1$ and $q_2$ are excited and $U(1)_F$ acts
non-trivially. Thus the moduli space for a single domain wall is
\be {\cal M}_1\cong {\bf R}\times {\bf S}^1 \ee
where ${\bf R}$ labels the center of mass while ${\bf S}^1$ labels
the phase.

\para
The low-energy dynamics of the wall is determined by allowing each
of the collective coordinates to vary over the worldvolume. Up to
two derivative terms, the dynamics is described by
\be {\cal L}_{\rm wall}=\frac{1}{2}T_{\rm wall}\int d^3x\
\,\partial_\mu X\partial^\mu X + \frac{1}{\Delta m^2}\
\partial_\mu\theta\,\partial^\mu\theta\ + \ {\rm fermions}
\label{dw1}\ee
where $\mu=0,1,2$ labels the domain wall worldvolume and $\Delta m
= (m_2-m_1)$. The periodicity is $\theta\in [0,2\pi)$. The fermion
zero modes are dictated by the fact that the domain wall is
$1/2$-BPS and the worldvolume theory therefore preserves four
supercharges, or ${\cal N}=2$ supersymmetry in three dimensions.
They are neutral and do not couple to $X$ or $\theta$ at the
two-derivative level.

\para
The periodic scalar $\theta$ on the worldvolume may be exchanged
in favor of a $U(1)$ gauge field $F_{\mu\nu}$ by the usual duality
construction,
\be 4\pi T_{\rm wall}\,
\partial_\mu\theta=\epsilon_{\mu\nu\rho}F^{\nu\rho}\label{dual}\ee
The normalization is fixed by the requirement that $\int {}^\star
dF\in 4\pi{\bf Z}$ as dictated by Dirac. This allows us to rewrite
the theory on the domain wall as a free ${\cal N}=2$ $U(1)$ gauge
theory in $d=2+1$ dimensions,
\be {\cal L}_{\rm
wall}=\frac{1}{4g^2}F_{\mu\nu}F^{\mu\nu}+\frac{1}{2g^2}\,\partial_\mu\psi
\,\partial^\mu\psi\ +\ {\rm fermions}\label{gaugeone}\ee
where the 3d gauge coupling is given by
\be \frac{1}{g^2}=\frac{T_{\rm wall}}{4\pi^2v^4}\ee
The scalar field $\psi$ in \eqn{gaugeone} has been canonically
normalized in a manner that befits the superpartner of the gauge
field. It is related to the center of mass $X$ by the scaling
$\psi=2\pi v^2X=T_{\rm vortex} X$. This normalization is the first
hint that the domain wall knows about the vortex.

\para
The simplest way to see that the domain wall is a D-brane for the
vortex string is to return to the original $X$ and $\theta$
collective coordinates in which we may find a "BIon"
spike\footnote{The full dynamics of the domain wall is governed by
the Born-Infeld action with the truncation to two derivatives
given by \eqn{gaugeone}. The BIon spike \eqn{bion} has the
property that it remains a solution of the full non-linear
Born-Infeld theory \cite{dbrane}.} solution on the worldvolume,
describing a string emanating from the domain wall
\cite{calmal,bion}. This BIon solution preserves 1/2 of the
supersymmetries on the wall and is given by,
\be \Delta m\,X+i\theta=\log \Delta m(x^1+ix^2)\label{bion}\ee
Since $\theta\rightarrow \theta+2\pi$ as we wind once around the
origin of the string at $x^1=x^2=0$, the dualization \eqn{dual}
ensures that the end of the string gives rise to a radial electric
field on the worldvolume: $F_{0r}\neq 0$. This means the end of
the string is electrically charged under the $U(1)$ gauge field on
the wall.

\para
The above discussion is from the perspective of the domain wall
worldvolume. We can also describe the vortex string ending on the
domain wall from the bulk perspective. The Bogomolnyi equations
describing this $1/4$-BPS configuration are a combination of the
domain wall equations \eqn{dw} and the vortex equations
\eqn{vortex} and were first derived in \cite{sy1}
\be \partial_1\phi=B_1\ \  ,\ \  \partial_2\phi=B_2\ \  ,\
\ \partial_3\phi=B_3-e^2(\sum_{i=1}^{N_f}|q_i|^2-v^2)\nn\\
{\cal D}_1q_i=i{\cal D}_2q_i\ \ ,\ \  {\cal D}_3q_i=(\phi-m_i)q_i\
\ \ \ \ \ \ \ \ \ \ \ \ \ \ \ee
Analytic solutions describing a string ending on the domain wall
are known only in the $e^2\rightarrow\infty$ limit
\cite{dbrane,isozumi}. At finite $e^2$ there is a binding energy
between the string and the domain wall \cite{mesakai} (see also
\cite{sybooj}). This finite, negative contribution to the total
energy scales as ${\cal E}_{\rm binding}\sim -\Delta m/e^2$. In
what follows we shall discuss the four-dimensional theory in the
$e^2\rightarrow \infty$ limit in which this binding energy
evaporates.

\section{D-Brane Interactions}

In this section we study the interaction of two or more domain
walls. Let's begin by reviewing what's known about the classical
scattering of domain walls, akin to bulk or ``closed string"
calculations in string theory.

\para
To study the scattering of multiple domain walls, we need to look
at a theory with $N_f\geq 3$ vacua. We start with $N_f=3$ and
study the system of domain walls interpolating between the vacuum
$\phi=m_1$ at $x^3\rightarrow -\infty$ and the vacuum $\phi=m_3$
as $x^3\rightarrow +\infty$. This system of domain walls was
studied in \cite{multiwall,mewall}. The general solution can be
thought of as two domain walls, separated by a modulus $R$. The
left hand wall has tension $T_1=(m_2-m_1)v^2$, while the right
hand wall has tension $T_2=(m_3-m_2)v^2$. In between the two
walls, the fields approximate the vacuum $\phi=m_2$. In addition
to their position collective coordinate, each wall also carries an
independent phase collective coordinate, both Goldstone modes
arising from the $U(1)^2_F$ flavor symmetry of the theory.

\para
In the moduli space approximation, the low-energy dynamics of
domain walls is described by the  $d=2+1$ dimensional, ${\cal
N}=2$ sigma-model with target space given by the domain wall
moduli space which is known to  be \cite{multiwall}
\begin{figure}[htb]
\begin{center}
\epsfxsize=3.2in\leavevmode\epsfbox{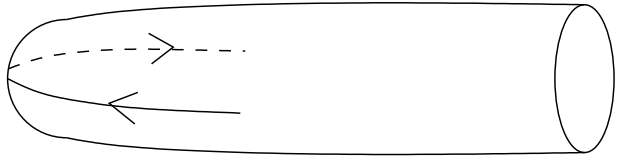}
\end{center}
\caption{The relative moduli space of two domain walls is a
cigar.}
\end{figure}
\be {\cal M}_{2}={\bf R}\times\frac{{\bf R}\times \tilde{\cal
M}_{\rm cigar}}{{\cal G}} \nn\ee
Here the first two ${\bf R}$ factors parameterize the
center-of-mass and overall phase of the soliton respectively. All
interesting dynamics is encoded in $\tilde{\cal M}_{\rm cigar}$,
the two-dimensional relative kink moduli space. The quotient by
the discrete group ${\cal G}$ acts only on the ${\bf S}^1$ fiber
of the cigar. For generic domain wall tensions, ${\cal G}={\bf
Z}$. However, when $T_1=T_2$, the second ${\bf R}$ factor
collapses to ${\bf S}^1$, and ${\cal G}\cong{\bf Z}_2$.

\para
The metric on ${\cal M}_{\rm cigar}$ was computed analytically in
\cite{mewall} in the $e^2\rightarrow\infty$ limit. (It does not
differ substantially from the metric at finite $e^2$ which was
computed numerically in \cite{sakmet}). The metric is smooth at
the origin and asymptotically, as $R\rightarrow\infty$, differs
from the flat metric on the cylinder by exponentially suppressed
corrections. However, the metric is not needed to determine the
crude features of domain walls scattering. Motion in the moduli
space which rounds the tip of the cigar (denoted in the figure)
corresponds to two approaching domain walls
which collide in finite time and rebound with their phases
exchanged. In particular, the cigar moduli space captures the most
important feature of the domain wall dynamics: the walls cannot
pass.

\para
Note that although the final result for the domain wall dynamics
is expressed in terms of a worldvolume theory, this is very much a
"bulk" calculation, with the moduli space ${\cal M}_2$ simply
encoding the interactions of the bulk fields in a concise
geometrical form.

\subsection{The Open String Description}

We will now describe a different realization of the domain wall
dynamics that doesn't involve solutions to the bulk field
equations, but instead arises by integrating in open strings
stretched between the two domain walls.

\para
The open string description starts by considering two, free,
domain walls with tensions $T_1$ and $T_2$. Their low-energy
dynamics is governed by two copies of \eqn{gaugeone}
\be {\cal L} = \sum_{a=1}^2\
\frac{1}{4g_a^2}\,F^{(a)}_{\mu\nu}F^{(a)\,\mu\nu} +
\frac{1}{2g_a^2}\,\partial_\mu\psi^{(a)}\partial^\mu\psi^{(a)}\ \
+\ {\rm fermions} \ee
where $1/g_a^2=T_a/4\pi^2 v^4$. Interactions between the domain
walls arise from the stretched, light vortex strings. Usually, one
shouldn't attempt to integrate in solitonic states in a field
theory; we shall discuss the regime of validity of this procedure
shortly. For now, let us ask what state the zero mode of the
string gives rise to in the worldvolume theory. In principle one
could quantize the theory on the vortex string with suitable
Dirichlet boundary conditions. (For example, a gauged linear sigma
model description of the vortex string theory was presented in
\cite{meami} to which the techniques of \cite{hiv} can be
applied). Here we determine the open string state using simple
consistency arguments.

\para
From the discussion of the previous section, we know that the open
vortex string has charge $(-1,+1)$ under the gauge fields
$(A_\mu^{(1)},A_\mu^{(2)})$ and gives rise to a state with mass
$T_{\rm vortex}(X^{(2)}-X^{(1)})=\psi^{(2)}-\psi^{(1)}$. Since
this state is BPS, preserving $1/2$ of the four supercharges on
the wall, it should come in short representation of supersymmetry.
This may be either a vector multiplet or a chiral multiplet. The
former is familiar from type II string theory when identical
D-branes approach, giving rise to non-abelian symmetry
enhancement. But, in our case, the D-branes are not identical:
they have different tension and carry different topological
charge. In such cases it is usual for the open string to give rise
to a chiral multiplet and I claim this indeed occurs for our
domain walls. Note that the virtual interactions of these states
must have a rather drastic effect, for they must stop the two
domain walls from passing each other. We will now see how this
comes about.

\para
The chiral multiplet consists of complex scalar $q$ and a Dirac
fermion $\lambda$, each with charge one under
$A_\mu=A_\mu^{(2)}-A_\mu^{(1)}$. Supersymmetry ensures that they
also couple to the separation $\psi=\psi^{(2)}-\psi^{(1)}$ and, in
particular, there is a Yukawa coupling on the wall
\be {\cal L}_{\rm Yuk} = \bar{\lambda}\psi\lambda \label{yuk}\ee
This ensures that when the walls are separated with $\psi>0$, the
fermion $\lambda$ gets a mass. However, as is well known,
integrating out a charged, massive, Dirac fermion induces a
Chern-Simons term for the gauge field \cite{redlich,avwitten}.
Similarly, integrating in such a fermion also induces a
Chern-Simons term. The low-energy energy effective action for the
relative motion of the domain walls is therefore given by the
Chern-Simons-Higgs theory, with action
\be {\cal
L}&=&\frac{1}{4g^2}\,F_{\mu\nu}F^{\mu\nu}+\frac{1}{2g^2}\,\partial_\mu
\psi\partial^\mu\psi +{\cal D}_\mu q^\dagger {\cal D}^\mu q +
\kappa\epsilon_{\mu\nu\rho}A^\mu F^{\nu\rho}\nn\\ &&\ \ \ \ \
-\psi^2|q|^2- \frac{g^2}{2}(|q|^2-\kappa \psi)^2 +\ {\rm
fermions}\label{lrel}\ee
where the fermionic terms are dictated by supersymmetry and
include, among others, the Yukawa coupling \eqn{yuk}. The
"reduced" coupling is given by $1/g^2=1/(g_1^2+g_2^2)$. Most
importantly, the Chern-Simons coupling is $\kappa=-1/2$. It arises
from integrating in the fermion $\lambda$ with mass $\psi>0$.
Notice the appearance of the $\kappa\psi$ coupling in the D-term:
this is the supersymmetric completion of the Chern-Simons
coupling.

\para
The lagrangian \eqn{lrel} is the open string description of the
relative domain wall dynamics. We wish to study the massless
degrees of freedom. However, at first glance it appears that the
separation $\psi$ of the D-branes is gapped, with the Chern-Simons
coupling giving both the photon and $\psi$  a mass $M\sim \kappa
g^2$, resulting in a unique classical vacuum of the theory at
$\psi=q=0$. Quantum effects change this conclusion. Suppose we
separate the D-branes by turning on $\psi\neq 0$. There is a
classical tadpole since the potential energy is
$V=g^2\kappa^2\psi^2/2$. The Chern-Simons coupling $\kappa$ is
renormalized at one-loop by integrating out the Dirac fermion
$\lambda$. This is given by \cite{redlich,avwitten}
\be \kappa_{\rm eff}=-\frac{1}{2}+{\rm sign}({\rm
Mass}[\lambda])=-\frac{1}{2}+{\rm
sign}(\psi)=\left\{\begin{array}{lr} 0 & \psi > 0 \\ -1 & \psi <0
\end{array}\right. \ee
When $\psi>0$, integrating out the fermion is simply the reverse
of the integrating-in procedure we performed to derive the
effective action \eqn{lrel}. Here the effective Chern-Simons
coupling vanishes and both the photon $A_\mu$ and $\psi$ are
massless, reflecting the fact that the D-branes are free to move.
However, things are very different when $\psi<0$. Here the
$\kappa_{\rm eff}\neq 0$ and there are no massless degrees of
freedom: the D-branes cannot move into this regime. We see that
the open string mode indeed prevents the domain walls from
passing.

\para
This method of using Chern-Simons interactions to lift portions of
the vacuum moduli space was previously studied in
\cite{menick,mecs}, following earlier work on the dynamics of 3d
${\cal N}=2$ theories \cite{deboer,ahiss}. Integrating out the
chiral multiplet also generates interactions between the massless
fields $\psi$ and $A_\mu$ on the wall worldvolume. Dualizing the
photon back to the relative phase $\theta$, the two derivative
terms in the effective Lagrangian are given by
\be {\cal L}=\ft12\,H(\psi)\ \partial_\mu\psi\,\partial^\mu\psi
+\ft12\,H(\psi)^{-1}\
\partial_\mu\theta\,\partial^\mu\theta\label{open}\ee
where $H(\psi)=1/g^2$ classically which, at one-loop, is corrected
to
\be H(\psi)=\frac{1}{g^2}+\frac{1}{2|\psi|}\label{cigar}\ee
This is the metric on a cigar. As $\psi\rightarrow\infty$, the
metric is that of a flat cylinder, up to power-law corrections.
This is in contrast to the bulk calculation, where the metric
differs by exponential corrections. This is not surprising: as we
review shortly, the calculations are valid in different regimes
and the K\"ahler potential receives no protection in theories with
four supercharges. Nevertheless, the basic features of the domain
wall dynamics are correctly captured by the open string
calculation. Note that the metric \eqn{cigar} is smooth at the
origin $\psi=0$, but cannot be trusted there since higher loop
corrections become important. Nevertheless, the results of
\cite{ahiss,mecs} guarantee that the true metric remains smooth at
the origin.

\subsection{Multiple Domain Walls}

We may repeat the above procedure for the $N_f-1$ walls in the
$U(1)$ theory with $N_f$ matter fields, interpolating between the
two outermost vacua $\phi=m_1$ and $\phi=m_{N_f}$. Each of the
$N_f-1$ walls has a different tension, $T_k=(m_{k+1}-m_k)v^2$. The
moduli space of domain wall solutions was studied in
\cite{multiwall,mewall}. Once again, solutions exist with
arbitrary separations between the walls. However, the ordering of
the walls is completely fixed, with $X^{(k+1)}\geq X^{(k)}$ where
$X^{(k)}$ is the position of the $k^{\rm th}$ domain wall. The
moduli space metric remains smooth as domain walls coincide.

\para
The open string description arises by integrating in chiral
multiplets between pairs of neighboring domain walls, each
generating a Chern-Simons term. The theory on the domain wall
worldvolume is therefore given by 3d ${\cal N}=2$
Maxwell-Chern-Simons-Higgs theory, with gauge group
$\prod_{k=1}^{N_f-1}U(1)_k$ and Chern-Simons couplings
\be -\frac{1}{2}\sum_{k=1}^{N_f-2} \,(A^{(k+1)}-A^{(k)})\wedge
(F^{(k+1)}-F^{(k)}) \ee
together with $N_f-2$ chiral multiplets, with charge $-1$ under
$U(1)_k$ and charge $+1$ under $U(1)_{k+1}$ for
$k=1,\ldots,N_f-2$. The fermions in these chiral multiplets have
mass $M_k=(\psi^{(k+1)}-\psi^{(k)})$. Upon integrating out the
fermions, the Chern-Simons couplings vanish only in the regime
$\psi^{(k+1)}>\psi^{(k)}$, mimicking the correct ordering of the
domain walls.

\subsection{Regimes of Validity}

Although the two computations described above lead to the same
crude physics of domain wall scattering, namely their ordering,
the details disagree. Most notably, the leading order velocity
interactions are exponentially suppressed in domain wall
separation in the bulk computation, while they are only suppressed
by power-law in the open-string computation. We shall now examine
the regime of validity of these two different approaches.

\para
The bulk calculation was classical, with the scattering of domain
walls determined by the field equations. The sigma-model in four
dimensions is not renormalizable and new UV degrees of freedom
must be introduced, giving expected corrections to the kinetic
terms of the form,
\be |{\cal D}q|^2\left(1 + \frac{1}{v^2}|{\cal
D}q|^2+\ldots\right) \ee
We insist that these corrections are negligible. The domain wall
interpolating between the $i^{\rm th}$ and $j^{\rm th}$ vacua has
width $\Delta m= |m_i-m_j|$, and these higher derivative
corrections may be ignored when:
\be \mbox{Bulk Criterion:} \ \ \ \frac{\Delta m}{v}\ll 1\ee
For the open string description to be valid, we require that the
stretched vortex string, which has mass $M\sim v^2 R$, is the
lightest degree of freedom. Furthermore, excited states of the
string should decouple. For the vortex strings, there are two such
excitations: oscillations have energy ${\cal E}_{\rm osc}\sim v$,
while internal excitations have energy ${\cal E}_{\rm int}\sim
\Delta m$. If we further impose that the separation $R$ between
the D-branes is greater than their width $L_{\rm wall}=1/\Delta
m$, then we learn that the open string description is valid in the
opposite regime,
\be \mbox{Open String Criterion:}\ \ \ \frac{v}{\Delta m}\ll 1\ee
so that the string oscillator states are lighter than the internal
modes. Note that it follows that the open string description is
only valid when $R\ll 1/v=1/T_{\rm vortex}^{1/2}$, which we may
call "sub-stringy" distance scales.

\subsubsection*{\it String Coupling}

Since the domain walls are D-branes for the vortex string, it is
natural to ask if they have the typical $1/g_s$ tension formula
familiar from type II string theory. Defining the inverse string
tension $\alpha^\prime = 2\pi/T_{\rm vortex}$, we can write
\be T_{\rm wall}=v^2 \Delta m =  \frac{\Delta
m}{v}\,\frac{1}{\alpha^{\prime\,3/2}}\ee
which suggests that the string coupling constant $g_s$ may be
\be g_s=\frac{v}{\Delta m}\label{coupling}\ee
I do not know if the ratio $v/\Delta m$ can indeed be interpreted
as the coupling of the vortex string. For small $e^2$, the
vortices are semi-classical objects which are known to reconnect
with probability one \cite{reconnect}: at weak gauge coupling, the
vortex strings are not weakly coupled. Nonetheless, as the
coupling is increased, this semi-classical treatment is no longer
valid and the vortices may potentially tunnel past each other. In
the sigma-model limit considered in this paper (defined by taking
$e^2\rightarrow\infty$ in the classical Lagrangian) the vortices
are infinitesimally thin and cannot be treated as semi-classical
objects. In this regime, a string coupling of the form
\eqn{coupling} is not implausible. It's interesting to note that
with this interpretation, the open string description is valid at
$g_s\ll 1$ as expected.

\subsection{A Further Generalization: Complex Masses}

So far we have described the moduli space of domain walls in the
theory with purely real masses $m_i$. In general these masses may
be complex and still be consistent with ${\cal N}=2$ supersymmetry
in four dimensions. Turning on a small, imaginary part for each
mass induces an attractive force between the domain walls. This is
most simply seen by studying the BPS tension formula, which now
reads
\be T_{\rm wall}=v^2|m_i-m_j| \label{comass}\ee
With $m_i-m_j$ complex, Pythagoras implies an attractive force
between two walls provided that a BPS bound state exists. From the
perspective of the moduli space approximation, this attractive
force can be described by a potential on the domain wall moduli
space which is proportional to the modulus of a Killing vector
arising from the $U(1)_F^{N_f-1}$ flavor symmetries. Complex
masses also allow for new $1/4$-BPS dyonic domain walls
\cite{dyonic} and domain wall junctions \cite{junction}.

\para
In the open string description, this attractive force arising from
complex masses is captured by a suitable FI parameter. This is
most simply described for the case of two walls. We may use the
$U(1)_R$ symmetry of the four dimensional theory to set ${\rm
Im}(m_3-m_1)=0$. Then the FI parameters are given by
\be \zeta = \frac{\left|\,{\rm Im}(m_2-m_1)\right|}{2\pi}=
\frac{\left|\,{\rm Im}(m_3-m_2)\right|}{2\pi}\ee
These appear in the scalar potential of the open string theory
\eqn{lrel} describing the domain wall dynamics, which now reads
\be V_{\rm wall}=\psi^2|q|^2+\frac{g_1^2+g_2^2}{2}
(|q|^2-\kappa\psi-\zeta)^2\ee
As the two walls are separated $\psi\rightarrow \infty$, the
potential is minimized by $|q|^2=0$ and flattens out to a constant
(recall that, after the one-loop correction, $\kappa=0$),
\be V_{\rm wall}\rightarrow \frac{(g_1^2+g_2^2)\,\zeta^2}{2}\ee
which, to leading order in $\zeta$, is equal to the binding energy
between the domain walls arising from \eqn{comass}. The unique,
quantum ground state of the theory is given by $\psi=0$ and
$|q|^2=\zeta$. This corresponds to the supersymmetric bound state
of two domain walls. The open string mode becomes tachyonic at
$\psi\sim g^2\zeta$ and condenses in the vacuum.

\section{D-Branes in Non-Abelian Theories}

So far we have discussed domain walls in abelian gauge theories
which have fixed ordering in space. We now turn to the more
general non-abelian theories where the ordering of domain walls is
more complicated. Nevertheless, we shall see that it is once again
captured by the quantum dynamics of a Chern-Simons theory on the
wall.

\para
We work with a four dimensional ${\cal N}=2$, $U(N_c)$ gauge
theory with $N_f$ matter multiplets in the fundamental
representation. As in section 2, we need only work with a subset
of the fields: a real adjoint scalar $\phi$ and $N_f$ complex
fundamental scalars $q_i$, $i=1,\ldots, N_f$. The scalar potential
is dictated by supersymmetry,
\be V= \sum_{i=1}^{N_f}q_i^\dagger(\phi-m_i)^2 q_i
+\frac{e^2}{2}\,\Tr\,(\sum_{i=1}^{N_f}q_i\otimes q_i^\dagger-v^2\,
1_N)^2\label{nalag}\ee
This potential has a large number of distinct, isolated, vacua.
There are supersymmetric vacua with $V=0$ only if $N_f\geq N_c$
and we assume this is the case. Each vacuum is determined by a
choice of $N_c$ distinct elements from a set of $N_f$
\be \Xi=\{\xi(a):\ \xi(a)\neq\xi(b)\ {\mbox {\rm for}\ } a\neq
b\}\ee
where $a=1,\ldots,N_c$ runs over the color index, and $\xi(a)\in
\{1,\ldots,N_f\}$. Choose $\xi(a)<\xi(a+1)$. Both terms in the
potential vanish by setting
\be \phi={\rm diag}(m_{\xi(1)},\ldots,m_{\xi(N_c)})\ \ \ \ , \ \ \
\ \ q^a_{\ i}=v\delta^a_{\ i=\xi(a)} \ee
The number of vacua of this type is
\be N_{\rm vac}=\left(\begin{array}{c}N_f \\
N_c\end{array}\right)=\frac{N_f!}{N_c!(N_f-N_c)!}\label{nvac}\ee
Each vacuum has a mass gap in which there are $N_c^2$ gauge bosons
with $M_\gamma^2=e^2v^2+|m_{\xi(a)}-m_{\xi(b)}|^2$, and
$N_c(N_f-N_c)$ quark fields with mass $M_q^2=|m_{\xi(a)}-m_i|^2$
with $i\notin\Xi$.

\para
Domain walls interpolate between a given vacuum $\Xi_-$ at
$x^3\rightarrow -\infty$ and a distinct vacuum $\Xi_+$ at
$x^3\rightarrow +\infty$. The first order non-abelian domain wall
equations are the obvious generalization of \eqn{dw},
\be {\cal D}_3\phi=-\frac{e^2}{2}(\sum_{i=1}^{N_f}q_i\otimes
q_i^\dagger-v^2\,1_N)\ \ \ \ ,\ \ \ \ {\cal
D}_3q_i=-(\phi-m_i)q_i\label{nadw}\ee
Solutions to these equations have tension given by,
\be T_{\rm wall}=
v^2[\Tr\phi]^{+\infty}_{-\infty}=v^2\sum_{i\in\,\Xi_+}m_i-v^2\sum_{i\in\,\Xi_-}m_i\ee
A strict classification of domain wall systems in this model
requires a specification of vacua $\Xi_-$ and $\Xi_+$ at left and
right infinity. However, this leads to a bewildering array of
possibilities, since the number of vacua \eqn{nvac} grows
exponentially with $N_f$. A coarser classification of domain wall
systems was introduced in \cite{mesakai} which contains the
important information about the topological sector, without
specifying the vacua completely. The first step is to introduce
the $N_f$-vector,
\be \vec{m}=(m_1,\ldots,m_{N_f})\ee
We can then write the tension of the domain wall as
\be T_{\rm wall}=v^2\,\vec{g}\cdot\vec{m}\ee
which defines a vector $\vec{g}$ that contains entries $0$ and
$\pm 1$ only. In analogy to the classification of monopoles
\cite{gno}, we further decompose this vector as
\be \vec{g}=\sum_{i=1}^{N_f}n_i\,\valpha_i\ee
where $n_i\in {\bf Z}$ and  $\valpha_i$ are the simple roots of
$su(N_f)$,
\be \valpha_1&=&(1,-1,0,\ldots,0)\nn\\
\valpha_2&=&(0,1,-1,\ldots,0)\\
\valpha_{N_f-1}&=&(0,\ldots,0,1,-1)\nn\ee
The condition that $\vg$ contains only $0$'s, $1$'s and $-1$'s
translates into the requirement that neighboring $n_i$'s differ by
at most one: $n_i=n_{i+1}$ or $n_i=n_{i+1}\pm 1$. Note that in
this notation, the domain walls in the abelian theory that we
discussed in section 3.2 have topological charge
$\vec{g}=(1,0,\ldots,0,1)$, or $n_i=1$ for all $i$.

\subsection{The Ordering of Domain Walls}

As in section 3, the bulk description of the domain wall dynamics
is studied in the moduli space approximation by examining
solutions to the static equations \eqn{nadw}. Given two vacua
$\Xi_-$ and $\Xi_+$ at left and right infinity, the number of
collective coordinates of the interpolating domain wall system is
determined by the vector $\vec{g}$ \cite{isozumi,mesakai}
\be \mbox{\# collective coords} = 2\sum_{i=1}^{N_f-1} n_i\ee
The interpretation of this result is that there are $N_f-1$ types
of "elementary" domain walls associated to the simple roots
$\vec{g}=\valpha_i$. The tension of the $i^{\rm th}$ elementary
domain wall is given by
\be T_i=v^2(m_{i+1}-m_i)\label{natension}\ee
A domain wall in the sector $\vec{g}$ decomposes into
$N=\sum_in_i$ elementary domain walls, each with its own position
and phase collective coordinate.

\begin{figure}[htb]
\begin{center}
\epsfxsize=4.2in\leavevmode\epsfbox{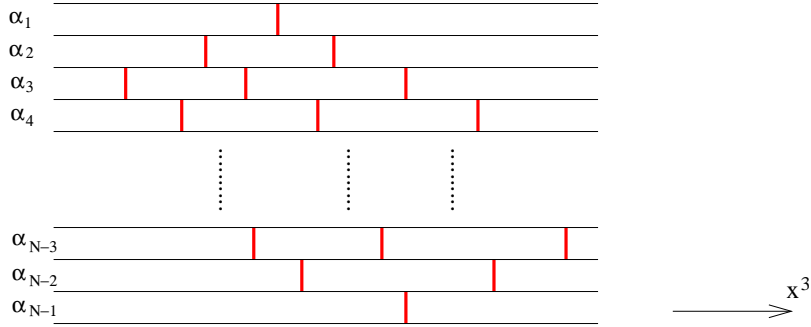}
\end{center}
\caption{The ordering of many domain walls. The horizontal
direction is their position, while the vertical denotes the type
of domain wall.} \label{interlace}
\end{figure}

\para
The moduli space of solutions was studied in some detail in
\cite{isozumi,nawalls}. Most important is the question of
ordering, for in the non-abelian theory it is no longer true that
walls can never pass through each other. Domain walls which live
in different parts of the flavor group, so that
$\valpha_i\cdot\valpha_j=0$, do not interact and  can happily move
through each other. When these domain walls are two of many in a
topological sector $\vec{g}$, an interesting pattern of interlaced
walls emerges, determined by which walls bump into each other and
which pass through each other. One finds \cite{nawalls} that the
$n_i$ elementary $\valpha_i$ walls must be interlaced with the
$\valpha_{i-1}$ and $\valpha_{i+1}$ walls. (The concept of
interlacing makes sense because $n_1=n_{i+1}$ or $n_i=n_{i+1}\pm
1$). The pattern of domain walls in space is depicted in figure
\ref{interlace} where $x^3$ is plotted horizontally and the
vertical position of the domain wall denotes its type. Notice that
the $\valpha_1$ domain wall is sandwiched between the two
$\valpha_2$ domain walls which, in turn, are trapped between the
three $\valpha_3$ domain walls. However, the relative positions of
the $\valpha_1$ domain wall and the middle $\valpha_3$ domain wall
are not fixed: these objects can pass through each other.

\para
This concludes the discussion of the bulk dynamics. It is rather
simple to reconstruct this ordering  from the open string
perspective. We start with the free $U(1)^N$ gauge theory on
$N=\sum_i n_i$ domain walls. The ordering described above arises
by integrating in chiral multiplets in the manner described in
section 3. The restriction that domain wall $i$ lies to the left
of domain wall $j$ may be enforced by the introduction of a chiral
multiplet with charge $(-1,+1)$ under the two gauge groups.
Integrating in this chiral multiplet gives rise once again to a
Chern-Simons term of the the form
\be -\frac{1}{2}(A^{(j)}-A^{(i)})\wedge (F^{(j)}-F^{(i}))\ee
and the associated supersymmetric completion in the D-term. The
arguments of the previous section generalize trivially. However,
there's a small subtlety that does not arise in the case of
abelian domain walls, since the $i^{\rm th}$ and $j^{\rm th}$
walls need not necessarily be neighbours. For example, the
$\valpha_1$ domain wall in the figure must must have two chiral
multiplets, one charged under each of the $\valpha_2$ gauge
fields. It doesn't interact directly with the $\valpha_3$ wall.
But only one of the $\valpha_2$ walls is a neighbour of the
$\valpha_1$ domain wall; on the other side sits the $\valpha_3$
domain wall. One may worry that no BPS string can pass through the
$\valpha_3$ domain wall, to end on the more distant $\valpha_2$
wall. In \cite{mesakai} a detailed study was performed on which
strings can end on which walls, and which strings may pass through
walls unaffected. One may check that there is always a BPS string
connecting a $\valpha_i$ domain wall with the two closest
$\valpha_{i+1}$ and $\valpha_{i-1}$ domain walls, consistent with
the conjecture of a chiral multiplet arising in from the open
string.

\para
There's one further difference with domain walls in the
non-abelian theory: we now have domain walls with the same tension
\eqn{natension} that may be possibly be classified as "identical".
Could strings stretched between the $\valpha_i$ domain walls give
rise to a non-abelian symmetry enhancement: $U(1)^{n_i}\rightarrow
U(n_i)$? To determine this really requires a semi-classical
quantization of the vortex string with Dirichlet boundary
conditions, but it appears that non-abelian symmetry enhancement
does not occur for these domain walls\footnote{An attempt at
non-abelian symmetry enhancement on the domain wall worldvolume
was made in \cite{sy2}, although not through the use of open
strings. The authors consider degenerate masses, resulting in a
non-abelian global symmetry on the wall. At the linearised level,
this may be dualised for a gauge symmetry, but it is unclear if
this gives rise to a non-abelian gauge symmetry at the non-linear
level.}. The key point is that identical walls are never
neighbours. A string connecting the walls only exists (at finite
$e^2$) if it is threaded with a confined monopole
\cite{memono,mesakai}. One may check, using the rules described in
\cite{mesakai}, that these strings are not BPS for all possible
positions of the intervening walls.

\para
To summarize: the open string theory on the walls in topological
sector ${\vec g}$ is given by $U(1)^{\sum_in_i}$ gauge theory,
with chiral multiplets linking the closest $\valpha_i$ and
$\valpha_{i\pm 1}$ walls, together with the associated
Chern-Simons interactions.

\section{Discussion}

In this paper we have developed the open string description of
semi-classical domain walls in a four-dimensional field theory. It
is instructive to compare the results against other field
theoretic D-branes\footnote{Other analogies between field
theoretic solitons and D-branes have been discussed in
\cite{dv,gomez}.}. As mentioned in the introduction, there are at
least two other examples of D-branes in theories without gravity
for which an open string description is known. In both cases, the
objects are D2-branes. Let's briefly review:

\subsubsection*{\it Monopoles in Little String Theory}

${\cal N}=(1,1)$ super Yang-Mills in six dimensions can be thought
of as the low-energy limit of type iib little string theory. When
the theory lives on the Coulomb branch, the spectrum of solitons
includes a monopole 2-brane. This is a D-brane for the instanton
string \cite{hw,jan}.

\para
The open string description of the monopole dynamics has much in
common with the domain walls described in this paper. (See, for
example, \cite{mdw} for a recent account of the relationship
between monopoles and domain walls). The moduli space of a single
monopole is ${\bf R}^3\times {\bf S}^1$. Since the monopole has
three-dimensional worldvolume, the periodic scalar may be dualised
in favor of a $U(1)$ gauge field living on the brane. The
instanton string may terminate on the monopole 2-brane, where its
end is electrically charged under this gauge field.

\para
BPS monopoles in an $SU(N)$ gauge group are classified by a magnetic
charge vector $\vec{g}=\sum_in_i\valpha_i$, where $\valpha_i$ are
the simple roots of $su(N)$, and $n_i\in {\bf Z}^+$ \cite{gno}.
Integrating in the open instanton strings\footnote{To my
knowledge, there has been no explicit semi-classical computation
of the open string spectrum by endowing instanton strings with
Dirichlet boundary conditions.} stretched between monopoles
enhances the free ${\cal N}=4$, $U(1)^{\sum_in_i}$ gauge theory on
the worldvolume to $\prod_{i=1}^{N-1}U(n_i)$ with bi-fundamental
hypermultiplets transforming in the $(\bar{\bf n}_i,{\bf
n}_{i+1})$ representation. The quantum dynamics of these theories
were  explored in \cite{sw,ch,me1} where it was shown that the
Coulomb branch metric of the three dimensional gauge theory
coincides with the metric of the appropriate monopole moduli
space. In this case, the open string description agrees with the
closed string description up to the two-derivative level, despite
being valid in different regimes. As usual, this can be traced to
a non-renormalization theorem (essentially the lack of
deformations of the appropriate hyperK\"ahler metrics).

\subsubsection*{\it Domain Walls in ${\cal N}=1$ Super Yang-Mills}

Unlike the two previous examples, domain walls in ${\cal N}=1$
$SU(N)$ super Yang-Mills cannot be treated semi-classically.
Nonetheless, the fact that they are BPS objects \cite{ds} means
that  they provide one of the few handles on the strongly coupled,
infra-red regime of the theory. Using M-theory techniques, Witten
showed that the domain wall is a D-brane for the QCD string
\cite{witten}.

\para
The open string description of the dynamics of $k$ domain walls
was determined by Acharya and Vafa \cite{bobby} by performing a
geometric transition in IIA string theory. The low-energy dynamics of the walls
are described by a 3d $U(k)$ gauge theory with  ${\cal N}=1$
supersymmetry. There is an adjoint multiplet, describing the
separation of the walls, and a Chern-Simons term at level $N$. In
a recent impressive calculation, Armoni and Hollowood have shown
how the attractive force between domain walls is captured by this
open string description at two loops \cite{aditim}. Although the
open string description agrees qualitatively with the bulk
calculation, it differs in the details. For example, the leading
order force arising from integrating out the open string zero mode
is power-law in the separation, whereas the bulk calculation gives
rise to an exponentially suppressed force. This is rather similar
to what we found above, with the open string kinetic term
\eqn{open} deviating by power-law corrections from the flat
metric, rather than exponential corrections.

\subsubsection*{\it An Underlying Closed String Description?}

In each of these three examples, the quantum open-string
description captures the classical bulk dynamics of the solitons.
The degree to which the two quantitatively agree is determined by
the amount of supersymmetry and the associated non-renormalization
theorems.  Of course, such agreement is commonplace within string
theory and is often summarized by the familiar annulus diagram in
which a closed string propagating at tree level between the two
branes may be re-interpreted as a vacuum loop of an open string
stretched between the branes.

\EPSFIGURE{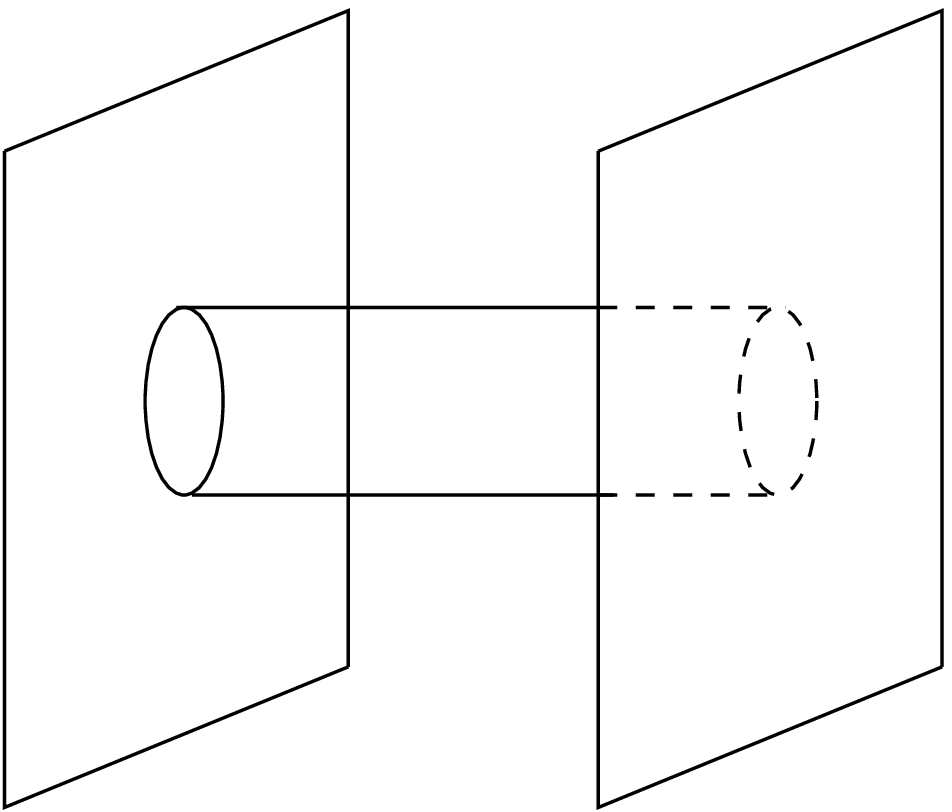,height=110pt}{}
\para
Both the four-dimensional and six-dimensional super Yang-Mills
theories described in this discussion have an underlying
non-critical string description. For the former, it is the usual
't Hooft expansion at large $N$; for the latter it is the type iib
little string theory propagating on the decoupled NS5-brane. In
each of these examples,  the bulk description of D-brane scattering
could well be called a ``closed-string" description. But what about
the ${\cal N}=2$ four-dimensional theory that is the main focus of
this paper? Is there also an underlying non-critical string
theory? The theory, when considered in the $e^2\rightarrow\infty$
limit, is a non-renormalizable massive sigma model. It is
interesting to speculate that there may be an underlying little
string theory of the vortex flux tubes providing the UV completion
of the theory. The existence of the open string description of the
worldvolume dynamics could then be interpreted in the familiar
framework of open-closed string duality.

\section*{Acknowledgement}
My thanks to Nick Dorey for many useful discussions. I'm supported
by the Royal Society.

\end{document}